\documentclass[APS,prl,twocolumn,superscriptaddress,amsmath,amssymb,preprintnumbers,floatfix,reprint]{revtex4-1}

\usepackage{graphicx,color,amsmath,amssymb}
\usepackage{slashed}
\usepackage{braket}
\usepackage{lipsum}
\usepackage{aas_macros}
\usepackage{physics}
\usepackage[export]{adjustbox}
\usepackage{tikz}
\usepackage{comment}
\usepackage{hyperref} % For hyperlinks in the PDF
\hypersetup{
    colorlinks=true,
    citecolor=blue,
    linkcolor=magenta,
    filecolor=magenta,
    urlcolor=blue,
}

\newcommand{\um}{{\, {\rm \mu m}}}

\newcommand{\eV}{{\, {\rm eV}}}

\newcommand{\gagg}{g_{a\gamma\gamma}}

\newcommand{\maxion}{m_a}
\newcommand{\g}{g_{a\gamma\gamma}}
\newcommand{\Deff}{D_{\rm eff}}
\newcommand{\thetaGC}{\theta_{\rm GC}}

\newcommand{\ie}{{\it i.e.~}}  \newcommand{\eg}{{\it e.g.~}}
\definecolor{mypurple}{RGB}{164,64,214}

\begin{document}

\title{Sensitivity of JWST to eV-Scale Decaying Axion Dark Matter}

% \author{People}
% \email{email}
% \affiliation{Places}

\author{Sandip Roy}
\email{sandiproy@princeton.edu}
\affiliation{Department of Physics, Princeton University, Princeton, NJ 08544, USA}

\author{Carlos Blanco}
\email{carlosblanco2718@princeton.edu}
\affiliation{Department of Physics, Princeton University, Princeton, NJ 08544, USA}
\affiliation{Stockholm University and The Oskar Klein Centre for Cosmoparticle Physics,  Alba Nova, 10691 Stockholm, Sweden}

\author{Christopher Dessert}
\email{cdessert@flatironinstitute.org}
\affiliation{Center for Computational Astrophysics, Flatiron Institute, New York, NY 10010, USA}

\author{Anirudh Prabhu}
\email{prabhu@princeton.edu}
\affiliation{Princeton Center for Theoretical Science, Princeton University, Princeton, NJ 08544, USA}

\author{Tea Temim}
\email{temim@astro.princeton.edu}
\affiliation{Department of Astrophysical Sciences, Princeton University, Princeton, NJ 08544, USA}

\date{\today}

\begin{abstract}

The recently-launched James Webb Space Telescope (JWST) can resolve eV-scale emission lines arising from dark matter (DM) decay. We forecast the end-of-mission sensitivity to the decay of axions, a leading DM candidate, in the Milky Way using the blank-sky observations expected during standard operations. Searching for unassociated emission lines will constrain axions in the mass range $0.18$ eV to $2.6$ eV with axion-photon couplings $\gagg \gtrsim 5.5 \times 10^{-12}$ GeV$^{-1}$. In particular, these results will constrain astrophobic QCD axions to masses $\lesssim$ 0.2 eV.

\end{abstract}

\maketitle

{\hypersetup{linkcolor=blue}}

\textbf{\textit{Introduction.---}} Pseudoscalar particles with feeble couplings to photons are ubiquitous in beyond the Standard Model constructions and are natural dark matter (DM) candidates. A famous example is the quantum chromodynamics (QCD) axion, which was originally proposed to resolve the Strong CP problem ~\cite{Peccei:1977ur,Peccei:1977hh,Weinberg:1977ma,Wilczek:1977pj}. Similar pseudoscalar axion-like particles (ALPs) arise independently in String Theory. For example, in ten-dimensional superstring theory ALPs arise as Kaluza-Klein modes in the compactification of $p$-form fields on 6-manifolds~\cite{Svrcek:2006yi,Arvanitaki:2009fg,Gendler:2023kjt}.  The non-detection of weak-scale dark matter (DM) candidates, strengthens the case that the missing $\sim 85\%$ of the matter of the universe is comprised of axions and/or ALPs. Therefore, the detection of pseudoscalars is pivotal if we hope to complete our cosmological and particle physics models of the universe. Here, we identify the recently launched James Webb Space Telescope (JWST) as a uniquely well-suited instrument to look for the astrophysical photon signatures of decaying axion and ALP DM. Note that for the remainder of this Letter, we refer to both QCD axions and ALPs as axions.

\begin{figure}[t!]
\includegraphics[trim = {1.5cm, .5cm, 1.5cm, .5cm},width=0.45\textwidth]{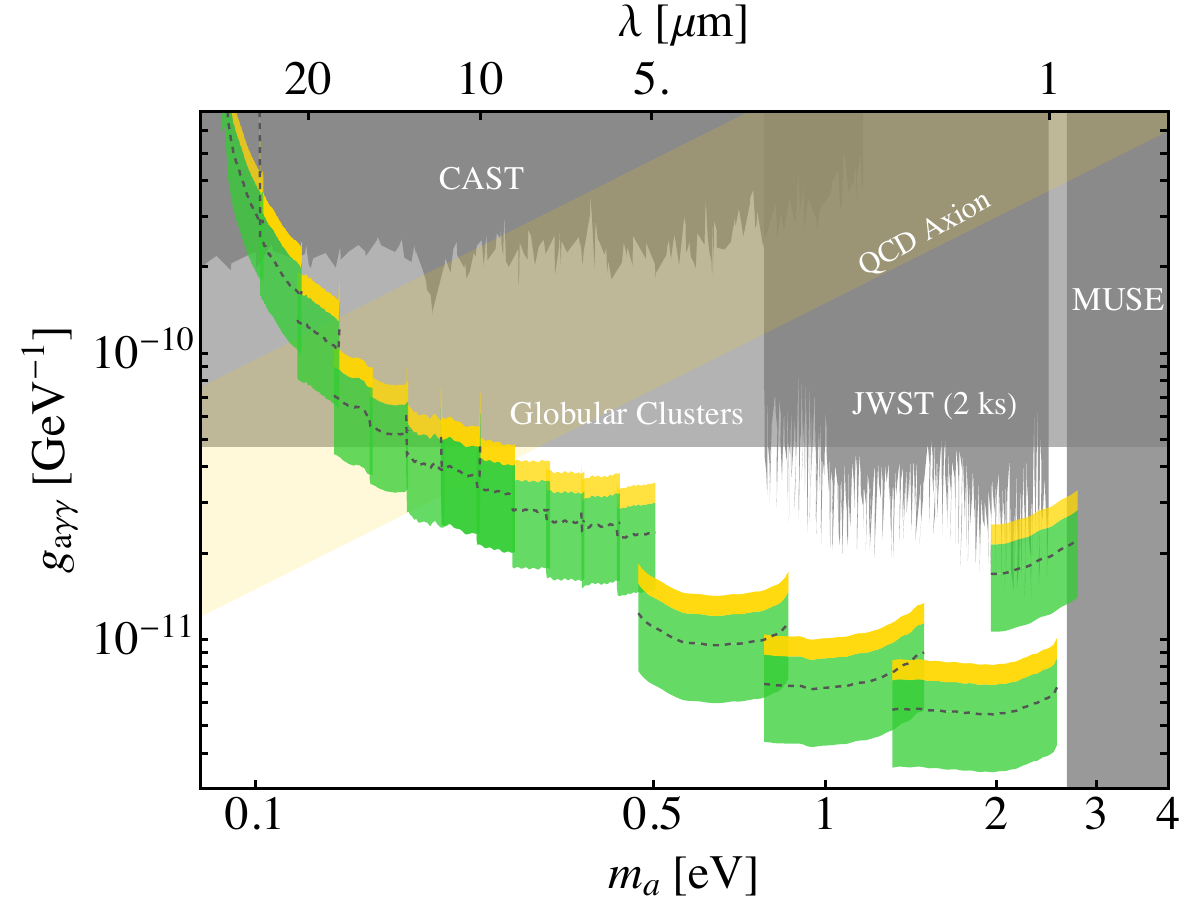}
\caption{
\label{fig:Sensitivity}
The expected 95\% upper limits (dotted lines) on the axion-photon coupling from an end-of-mission analysis of JWST observations. The solid green (yellow) region encloses our projections at 1(2)-$\sigma$ containment. We show in gray the existing constraints on axions~\cite{Dolan:2022kul,CAST:2008ixs,CAST:2013bqn,janish2023hunting,Todarello:2023hdk}. Possible QCD axion models live in the transparent yellow region. In order to consistently compare our projections with the recent JWST analysis~\cite{janish2023hunting}, we rescale their constraints according to the NFW profile adopted in this work.
}
\end{figure}

Axion cosmology is governed by the scale of symmetry breaking, $f_a$, that leads to the axion as a Goldstone boson and that of non-perturbative effects, $\Lambda$, that give the axion a periodic potential, such as $V(a) = \Lambda^4 \cos(a/f_a)$, where $\Lambda^4 = m_a^2 f_a^2$. If the symmetry is broken before inflation, the present-day axion abundance is determined by the misalignment mechanism~\cite{Irastorza_2018},
\begin{align} \label{eqn:abundance}
    {\Omega_a \over \Omega_{\rm DM}} \approx \left( {m_a \over {\rm eV}} \right)^{1/2} \left({f_a \over 10^{11} \ {\rm GeV}} \right)^2 \left( {a_i \over \pi f_a /\sqrt{3}} \right)^2,
\end{align}
where $a_i$ is the amplitude of the axion field at the time of symmetry breaking. Therefore, axions with non-thermal misalignment production mechanisms are consistent with a vast parameter space~\cite{Arias:2012az}, living anywhere in unexplored parameter space in Fig.~\ref{fig:Sensitivity}, motivating searches over a broad mass range. If the symmetry is broken after inflation ends, the axion field takes on uncorrelated values in distinct Hubble patches, leading to topological defects such as axion strings and domain walls that decay and provide contributions to the axion DM abundance~\cite{benabou2023signatures}. 
The QCD axion lives on the narrow band with a linear relationship between the QCD axion mass and its coupling to photons in Fig.~\ref{fig:Sensitivity}. Then the axion mass and potential are temperature-dependent, leading to a modification to Eq.~\eqref{eqn:abundance} in the pre-inflationary scenario~\cite{diCortona:2015ldu}. In the post-inflationary scenario, recent cosmological simulations of the topological defects find that the observed DM density is consistent with axion masses up to $m_a \gtrsim$ meV~\cite{Gorghetto:2020qws} (although, Refs.~\cite{Buschmann:2019icd, Buschmann:2021sdq} find masses in the range 40 $\mu$eV $\le m_a \le$ 180 $\mu$eV). 

In flavor-universal models, the QCD axion mass is constrained to be $m_a \lesssim 0.02$ eV by anomalous cooling bounds from neutron stars~\cite{Buschmann:2021juv} and supernovae~\cite{Lella:2023bfb}. However, cooling constraints can be relaxed for axions with nucleophobic couplings, e.g. flavor non-universal scenarios~\cite{DiLuzio:2017ogq, Bjorkeroth:2018ipq, Badziak:2023fsc}, opening up the possibility of eV-scale QCD axions. Indeed, there are many models in which the QCD axion DM mass is in the range meV $\lesssim m_a \lesssim$ eV~\cite{Co:2017mop, Caputo:2019wsd, Visinelli:2009kt, Co:2019jts, cyncynates2023heavy}.

Much of the effort to detect axions in the laboratory and in astrophysical settings relies on the axion-photon interaction $\mathcal{L} \supset -\gagg a(x) F \tilde{F}/4$, where $a(x)$ is the axion field, $\gagg = C_{a\gamma\gamma}\alpha/2\pi f_a$ is the axion-photon coupling constant, with $C_{a\gamma\gamma}$ an $\mathcal{O}(1)$ number, $F$ represents the electromagnetic field strength tensor, and $\tilde{F}$ is its dual. In the mass range of interest, model-independent constraints on $\gagg$ come from the CAST helioscope experiment~\cite{CAST:2008ixs, CAST:2013bqn} and stellar evolution in globular clusters~\cite{Dolan:2022kul}. In addition, laboratory experiments have been proposed to probe the same region of parameter space~\cite{IAXO:2019mpb, Liu_2022, Baryakhtar:2018doz}. Another powerful technique for detecting axions is to look for their decay products. In particular, axions can decay into photons with rate $\Gamma_a = \gagg^2 m_a^3/64\pi$~\cite{ParticleDataGroup:1996dqp}. Searches for photons from decaying axions spans from radio~\footnote{At low frequency the decay rate is Bose-enhanced relative to the previously-quoted vacuum decay rate.} to gamma rays~\cite{Caputo:2018vmy,Chan:2021gjl,Sun:2021oqp,Todarello:2023hdk,Grin:2006aw,Dessert:2018qih,Foster:2021ngm,Calore:2022pks,DeRocco:2022jyq,Blanco:2018esa}. Searching for photons from decaying axion DM in the range meV $\lesssim m_a \lesssim$ eV requires a highly sensitive infrared telescope, such as the JWST.

JWST represents a substantial leap in space-based imaging technology and is the scientific successor to the Hubble Space Telescope. Designed for infrared (IR) observation at the diffraction limit, JWST is capable of broad- and narrow-band imagery and integral-field spectroscopy in the wavelength range of $ 0.6 \; \mu \text{m} \leq \lambda \leq 29 \; \mu \text{m}$, which corresponds to photon energies between about 0.05 and 2 eV~\cite{greenhouse2016james}. JWST is equipped with instruments that are uniquely suited for the search for IR-scale decaying DM, namely, the Near-Infrared Spectrograph (NIRSpec) Integral Field Unit (IFU)~\cite{2022AA...661A..80J}, and the Mid-Infrared Instrument (MIRI) Medium Resolution Spectrometer (MRS)~\cite{2015PASP..127..646W}. 

In this Letter, we show that JWST will have leading sensitivity to eV-scale axion decay at the end-of-mission. Because the Earth is embedded in the Milky Way (MW) DM halo, every JWST observation ever taken is sensitive to axion decay. A particularly efficient approach for DM decay searches is to analyze blank-sky observations~\cite{Dessert:2018qih,Foster:2021ngm,Dessert:2023vyl}, which are observations of low-surface-brightness locations in the sky. In the Mikulski Archive for Space Telescopes (MAST) database~\cite{MAST}, we find 9.4 Ms of data usable for this purpose.

\begin{figure*}[t]
\includegraphics[trim = {1cm, 1cm, 0cm, 1cm},width=0.55\textwidth]{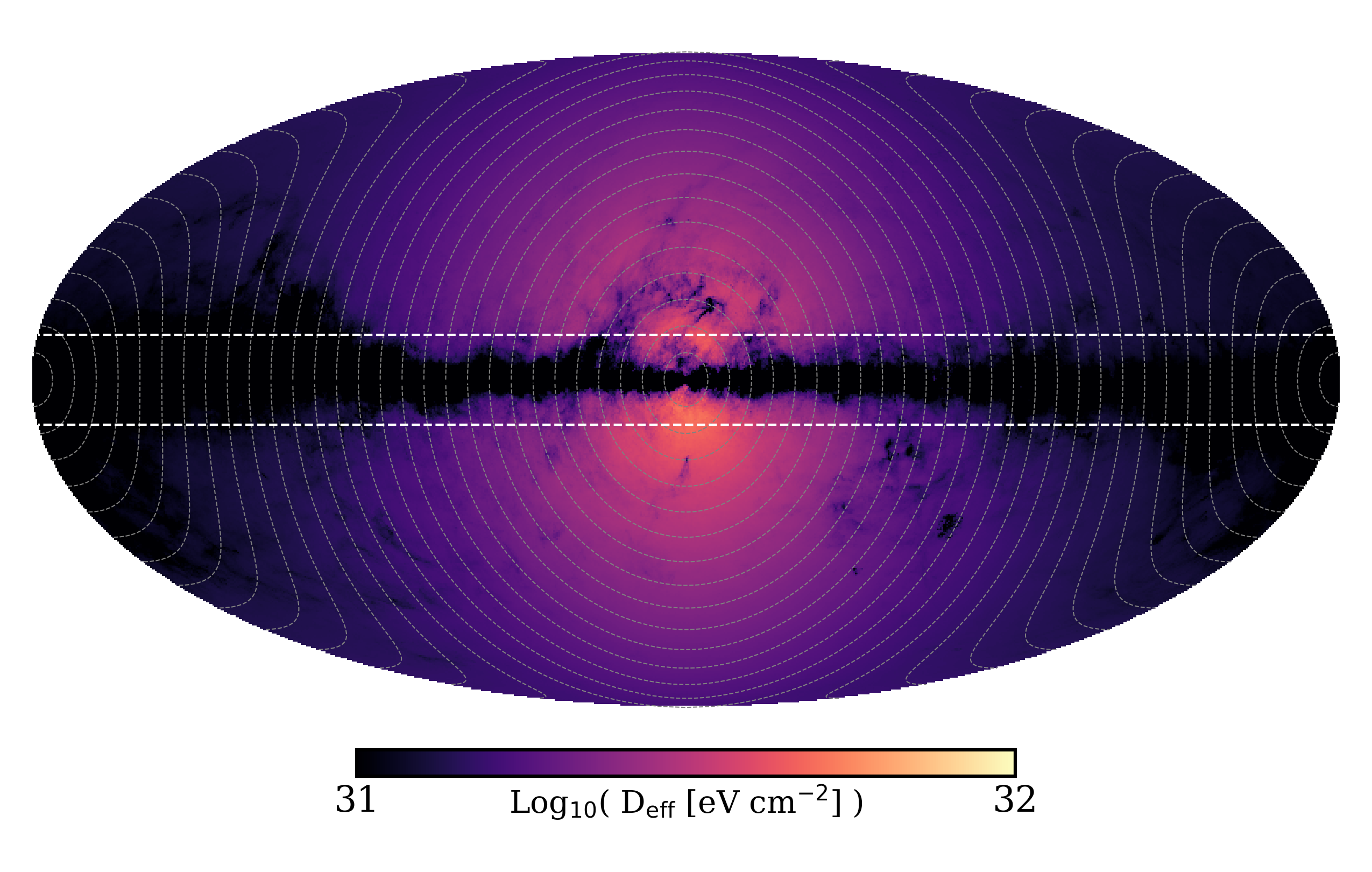}
\includegraphics[trim = {.5cm, .5cm, 1cm, 5cm},width=0.44\textwidth]{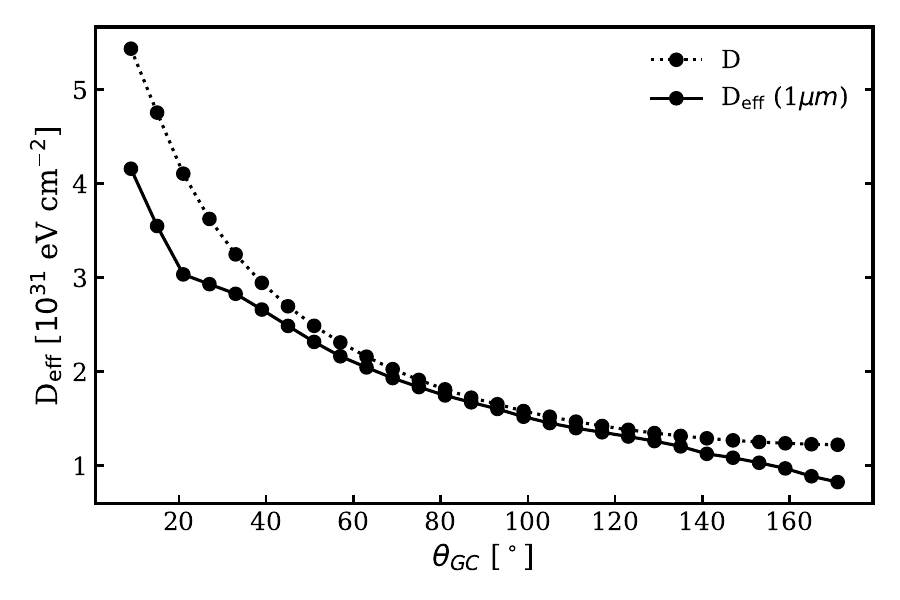}
\caption{
\label{fig:D-effective}
\emph{Left panel}: $D_{\rm eff}$ in galactic coordinates at $1\um$, corresponding to $m_a = 2.5 \eV$. The grey dashed lines delineate our ring edges and the dashed white lines are $|b|=10^\circ$, below which we mask out in our analysis.
\emph{Right panel}: The angular-averaged $D_{\rm eff}$ at a wavelength of $1\um$ (solid line) and without any dust extinction (dotted line), avoiding the masked galactic plane region. Lines of sight close to the galactic plane at $\thetaGC = 0^\circ$ and $\thetaGC=180^\circ$ experience the strongest dust extinction. At wavelengths greater than $1\um$, the dust extinction decreases as shown in detail in the Supplemental Material.
}
\end{figure*}
\vspace{1cm}
\textbf{\textit{Computing the Axion-Decay Photon Flux.---}}
The expected flux density from axion decays is expressed as~\cite{Lisanti:2017qoz}
\begin{equation}
\label{eq:axionflux}
    \dfrac{d\Phi}{d\lambda}(\lambda,l,b) = \dfrac{m_a}{2}\dfrac{\Gamma_a}{4\pi m_a} \dfrac{dN_\gamma(l,b)}{d\lambda} D(l,b)
\end{equation}
where $d\Phi/d\lambda$ has units of [MJy/sr Hz/$\um$], $dN_\gamma/d\lambda$ is the axion decay flux at Earth which depends on galactic coordinates $(l,b)$ through Doppler broadening, shifting and dust extinction, and the D-factor $D = \int_{\rm l.o.s.}ds \,\rho_a(ds,l,b)$ is the integrated DM density along the line of sight (l.o.s.) in units of [eV/cm$^2$]. Often the Doppler broadening and shifting of the DM particles can be ignored, but JWST has $\mathcal{O}(0.1\%)$ energy resolution, meaning that it can resolve emission lines produced by axion decay. Furthermore, lines of sight near the Galactic plane (GP) can be attenuated due to extinction by interstellar dust. Including all these effects, we have
\begin{equation}
    \label{eq:dNdE}
    \dfrac{dN_\gamma}{d\lambda} = \dfrac{m_a}{2}\dfrac{\int_{\rm l.o.s.} ds \exp(-\int_0^s ds' n_d \sigma) \rho_a(s) f(v(\lambda);r)}{\int_{\rm l.o.s.} ds\rho_a(s)} 
\end{equation}
where $f(v(\lambda);r)$ is the isotropic DM velocity distribution in~\cite{Dessert:2023vyl}. That velocity is a function of the observed wavelength via $|v(\lambda)|=2/(m_a\lambda(1-{\bf\hat{n}}\cdot {\bf v_\odot})) - 1$, where ${\bf\hat{n}}$ is the direction along the line of sight and ${\bf v_\odot}$ is the velocity of the Sun with respect to the galaxy. The extinction is provided by the exponential factor, where $n_d(s,l,b)$ is the dust density and $\sigma(\lambda)$ is its cross-section with photons. Although some 3-dimensional models of the galactic dust distribution exist~\cite{Green_2019}, they are typically not reliable out to distances of order the Milky Way scale radius, so in this work we conservatively model the extinction as if all emission originated at infinity, so that $\exp(-\int_0^s ds' n_d \sigma) \to 10^{-0.44A_\lambda}$ where $A_\lambda$ is the galactic extinction at wavelength $\lambda$. We use the galactic dust models in~\cite{1998ApJ...500..525S,doug_2011} and the extinction curve from \cite{Gao_2013} to compute the galactic extinction (for details, see the Supplemental Material (SM)).
The axion decays to a diphoton final state, meaning that $\int d\lambda dN_\gamma/d\lambda = 2\times 10^{-0.44A_{2/m_a}}$ and the total expected flux simplifies to $\Phi = \Gamma_a D_{\rm eff} / 4\pi$, where $D_{\rm eff}$ is the D-factor times the attenuation due to dust. We compute the D-factor in the Milky Way using a Navarro-Frenk-White (NFW) profile~\cite{Navarro:1995iw,Navarro:1996gj} with DM density at Earth $\rho_{\rm Earth} = 0.29$ GeV/cm$^3$ and a scale radius $r_s = 19.1$ kpc, using recent results calibrated on Gaia DR2 data~\cite{Cautun_2020}, along with a GC-Earth distance $d_{\rm Earth} = 8.23$ kpc~\cite{Leung_2022}. $\Deff$ at $1\um$ (\ie $m_a=2.5\eV$) is shown in Fig. \ref{fig:D-effective}. With knowledge of the astrophysical parameters, the axion DM decay flux is given by~\cite{Grin:2006aw} 
\begin{widetext}
\begin{equation}
\Phi(l,b) = 1.0\times10^{-9}\ \textrm{erg/cm}^2\textrm{/s/sr} 
\left(\dfrac{\g}{10^{-11}\ \textrm{GeV}^{-1}} \right)^2
\left(\dfrac{m_a}{1\ \textrm{eV}} \right)^3
\left(\dfrac{D_\textrm{eff}}{10^{31}\ \textrm{eV}/{\rm cm}^2} \right)
\end{equation}
\end{widetext}
where the line wavelength $\lambda = 2.5\ (1\ \textrm{eV}/m_a)\ \mu$m, corresponding to an energy $E_\gamma = m_a/2$. In particular, the MIRI MRS filters are sensitive to axions with masses $0.089$ eV $\leq m_a \leq 0.506$ eV, while the NIRSPEC IFU filters are sensitive to $0.47$ eV $\leq m_a \leq 2.75$ eV, so that our projections cover unexplored parameter space over an order of magnitude in axion mass.

In this work, we compute the expected sensitivity of JWST to axion decays in the Milky Way halo. To this end we need to know the total exposure time usable for blank-sky observations. Every IFU observation taken is useful, unless the source is sufficiently extended such that emission will dominate the observation. On the other hand, point sources can be removed via a spatial mask. JWST's lifetime is fuel-limited, and the telescope was designed to operate for 10 years, but is expected to continue operating for about 20 years. Not all of this time will be spent observing the blank sky; in particular, the observing efficiency is expected to be around 70\%~\cite{JWST_Docs}. We pay an additional penalty because only some fraction of this time will be spent observing with the IFU modes. We compute the expected exposure time in each instrument mode in a data-driven way by extrapolating the first $\sim$1.5 years of observations, taken between March 17, 2022 and Nov 6, 2023, to that expected in 10 years of operation. We obtain the exposure times from the MAST database. In this computation we exclude observations with extended sources and observations within the GP mask, which collectively account for approximately 30\% of the total exposure time. However, we note that extended source observations are typically accompanied by dedicated background observations, so that this approach is extremely conservative. Our expected exposure times are shown in Tab.~\ref{tab:exposure}. Given that the JWST IFU field of view (FOV), $\mathcal{O}(10\ \textrm{arcsec}^2)$ depending on the observing mode, is only 1 part in $10^{11}$ of the full sky at any time, we need to know the observing pattern. We assume that JWST observes isotropically across the full sky. In reality this distribution is likely to be biased towards the Galactic Center (GC) and Galactic Plane (GP), \ie towards regions of larger signal.

We use the \texttt{jwst\_backgrounds}~\footnote{\href{https://jwst-docs.stsci.edu/jwst-other-tools/jwst-backgrounds-tool}{https://jwst-docs.stsci.edu/jwst-other-tools/jwst-backgrounds-tool}} tool developed by JWST to compute the background contribution at a given point on the sky. The astrophysical background is modeled with four components. The dominant contributions are the in-field zodiacal light \cite{Kelsall_1998,Wright_1998,Gorjian_2000} and the in-field interstellar medium (ISM) emission \cite{1998ApJ...500..525S}. Zodiacal emission arises through the reflection of sunlight by dust in our solar system and thus dominates the background at low ecliptic latitudes. The ISM background is produced by dust emission within the Milky Way and thus dominates at low galactic latitudes. Other background components include the astrophysical and detector stray-light \cite{Lightsey2016,Wainscoat1992} and the thermal self-emission which dominate at wavelengths $\gtrsim 20\um$. For further details and a measurement of the astrophysical background, see \cite{Rigby_2023}. 

\textbf{\textit{Projecting JWST's Sensitivity to Axion Decays.---}}Using the procedure developed in Ref.~\cite{Foster:2021ngm}, we bin the data in 30 concentric rings around the GC of width $6^\circ$ and mask the GP up to $b \leq 10^\circ$ so that the innermost and outermost rings are entirely masked. This mask is motivated by that the dust extinction can be difficult to calculate in the GP due to a larger variety of dust molecules~\cite{Gao_2013}, and we wish to remain insensitive to such details. As we show in the SM, smaller masks actually reduce $D_{\rm eff}$ near the GP, although we would gain $\sim$10\% in exposure time. We compute the expected exposure time in each JWST spectroscopic mode and ring. Note that there are 12 MRS modes, but they are split into groups of three by wavelength, and the entire group is observed simultaneously. There are 8 NIRSpec filters over four wavelength bands. Each band is observed in high and medium resolution. We do not consider the NIRSpec PRISM mode, which has reduced spectral resolution.

The axion decay flux is modeled as in Eq.~\ref{eq:axionflux}, where $D_{\rm eff}$ is computed by averaging over each ring. In the course of this averaging, the line is broadened beyond the intrinsic width of 220 km/s of Doppler broadening because the Doppler shifting can change significantly over a ring. We could shift the data to the galactic rest frame, but this introduces difficult-to-model bin-to-bin correlations in the data~\cite{Dessert:2023fen}. We compute the root-mean-square Doppler shift in each ring and add it in quadrature to the Doppler broadening to determine the observed width of the line $v_{\rm line}$. For the innermost ring, nearly transverse to the solar motion, the width is 220 km/s; for the rings 90$^\circ$ from the GC the width increases to $\sim$275 km/s, (for details, see SM).

\begin{table*}[t]
\begin{center}
\tabcolsep=0.08cm
\begin{tabular}{cccc}
\textbf{MIRI MRS Grating} & Short & Medium & Long\\
\hline
$t_{\rm exp}$ [Ms] & 12.5 & 12.4 & 12.5 \\ 
\end{tabular}\\
\begin{tabular}{ccccccccc}
\textbf{NIRSpec} & F070LP & F100LP & F170LP & F290LP & F070LP & F100LP & F170LP & F290LP \vspace{-0.1cm} \\
\textbf{Filter} & G140M & G140M & G235M & G395M & G140H & G140H & G235H & G395H\\
\hline
$t_{\rm exp}$ [ks] & 1.4 & 750 & 1060 & 1980 & 5.9 & 1300 & 6900 & 8070\\ 
\end{tabular}
\caption{\label{tab:exposure} The exposure times $t_{\rm exp}$ in [Ms] for MIRI and [ks] for NIRSpec assumed in our projections for each observing mode.}
\end{center}
\end{table*}

To model the expected signal-to-noise ratio (SNR) for the axion decay, we use the output of the JWST Exposure Time Calculator (ETC)~\cite{ETC} in each analysis ring assuming a fixed exposure time $t_\textrm{exp}$ and axion-photon coupling $\gagg$. This approach is agnostic to any particular analysis strategy, such as parametric likelihood-based frequentist modeling as in Ref.~\cite{Dessert:2018qih} or nonparametric approaches such as Gaussian Processes as in~\cite{Frate:2017mai,Foster:2021ngm}, which should be developed when performing analysis on real data. However, we verified on simulated data that the parametric approach, where, \eg, the signal line is modeled along with a quadratic background model, returns results consistent with that of the ETC. Note, however, that we have not included possible systematics into this projection. For instance, we do not account for diffuse astrophysical line emission. In the vicinity of bright astrophysical lines, our limits may disappear; however, these lines exist over only a small fraction of the parameter space. There are known possible spurious line signals~\cite{JWST_Docs}: scattered light from a bright line emitters and the 12.2$\um$ MRS spectral leak. The former should affect only a few individual observations, and particular wavelengths with less-understood instrumental spectral features can be avoided, so we expect these issues to have minimal impact on our results.

The ETC incorporates sky-dependent background models; we query the background model at the point in the ring closest to the ecliptic plane where the backgrounds are largest outside $|b|\leq 10^\circ$. We show the effects of other choices in the SM. We query the ETC for each observing mode at an axion mass such that the decay occurs in the mode's wavelength range. The axion signal in each ring is input as a spectral line positioned at a frequency $\maxion/2$ with the observed ring-dependent width. We use the scaling relation in Eq.~\ref{eq:snr_scaling}, verified empirically with the JWST ETC, to determine the SNR across the entire filter bandwidth and repeat this process for each NIRSPEC and MIRI filter of interest (where the $N_a$ and $N_{\rm bkg}$ represent the detector counts for axions and background respectively, $\epsilon_\lambda$ represents the wavelength-dependent filter throughput and $\Phi_{\textrm{bkg},\lambda}$ represents the wavelength and line-of-sight-dependent background and instrumental noise flux). We then compute Asimov likelihoods~\cite{Cowan:2010js} in each ring and filter and multiply the likelihoods in each ring to obtain a joint likelihood. For each value of $m_a$, we solve for the 95\% upper limit on the axion-photon coupling strength $\g^{95}$ such that SNR$|_{\lambda=2/m_a} = \sqrt{2.71}$, given that Wilks' theorem holds. 

\begin{equation}
\label{eq:snr_scaling}
    \textrm{SNR}(\lambda) = \frac{N_a}{\sqrt{N_{\rm bkg}}} \propto m_a^3 \,\g^2\,D_{\rm eff}\,t_\textrm{exp}^{\frac{1}{2}}\epsilon_\lambda^{\frac{1}{2}} \Phi_{\textrm{bkg},\lambda}^{-\frac{1}{2}}.
\end{equation}

We present our projected 95\% upper limits in Fig.~\ref{fig:Sensitivity}, along with their 1- and 2-$\sigma$ enclosing regions. In particular, we show that JWST will have leading sensitivity to axions over an order of magnitude in mass $0.18$ eV $\leq m_a \leq 2.6$ eV. The NIRSpec IFU filters, covering axion masses between $0.5$ eV $\leq m_a \leq 2.6$ eV, reach axion-photon couplings down to $5.5 \times 10^{-12}$ GeV$^{-1}$. We show our MIRI MRS projections in the twelve bands at smaller masses, which are somewhat weaker, although they promise to probe astrophobic QCD axion scenarios down to $0.18$ eV. Our analysis shows that JWST has the ability to rule out a QCD axion DM candidate above about 0.2 eV, whether or not it is coupled to baryonic matter, making this analysis particularly complementary to other astrophysical probes.

\textbf{\textit{Discussion \& Conclusion---}}JWST is the first telescope with exquisite enough sensitivity to near- and mid-IR emission lines to probe viable axion DM. Although JWST was not designed with axion decay searches in mind, it is nevertheless a powerful tool in the search for axions owing to its spectral resolution, which is precise enough to resolve lines resulting from DM decay. In this work we show that an analysis of the end-of-mission blank-sky observations will cover novel axion parameter space in a region that is currently unexplored by terrestrial experiments. This will effectively rule out a QCD axion DM candidate heavier than about 0.2 eV, regardless of its matter couplings. Importantly, the blank-sky observations discussed in this work will be made in the course of the normal operations of JWST, and require no changes in the observing strategy. 

Another search strategy could be to observe a dwarf galaxy such as Draco dSph. Typically the blank-sky strategy is stronger, but the JWST FOV is significantly smaller than most telescopes used to search for DM decay, while the dwarf D-factors increase when averaged over smaller FOVs around their centers. Furthermore, dust extinction reduces the MW D-factor while it is irrelevant for dwarfs not in the plane of the galaxy. Accounting for these effects, Draco hosts a larger D-factor than any MW location, which we estimate to be $\sim (9 \pm 2.5) \times 10^{31}$ eV/cm$^2$ in the JWST FOV (the variation with observing mode is small)~\cite{Evans:2016xwx}, which is conservatively about three times as large as the MW average. The typical dwarf velocity dispersion is also smaller than the MW, $\mathcal{O}(10^{-4})$~\cite{2012AJ....144....4M}, which means that DM decay lines originating in dwarfs would be unresolved by JWST, but this has a relatively minor effect on our sensitivity. Draco and other dwarfs have not to-date been observed by JWST except in imaging modes, and are unlikely to be observed in normal operations. A typical single-target observation could achieve total exposure times of around 100 ks, about 200 times smaller than that of our combined exposure. Therefore, such an analysis could not exceed the sensitivity of an end-of-mission blank-sky analysis (except in the mass range from 2.55 to 2.66 eV, which is probed only by the low-exposure \texttt{F070LP} filters). For example, assuming a 100 ks observation of Draco across the four high-resolution NIRSpec modes, we estimate a peak sensitivity of $\gagg^{95} \approx 8\times 10^{-12}$ GeV$^{-1}$. However, a 100 ks dwarf observation in one particular filter would result in sensitivity over a smaller mass range which is competitive with that forecasted here.

In this work we focus on forecasting the JWST end-of-mission sensitivity to axion decay. JWST has now been in operation for $\sim$1.5 years, and therefore has collected 15\% of its total data. An analysis of only the currently-available data would still provide leading sensitivity, but weaker than that projected here by a factor $0.15^{-1/4} \sim 1.6$, which we leave to future work.

\textbf{Note added:} During the preparation of this manuscript, Ref.~\cite{janish2023hunting} appeared on the arXiv, also studying Milky Way axion decay signatures using JWST. Our works have some overlap but are complementary; the main result of that work is the analysis of two NIRSpec observations. While that work projects the end-of-mission sensitivity for two NIRSpec filters through extrapolation of their limits, we take into account spatial information in both the decay signal and the astrophysical backgrounds to project sensitivity for all JWST IFU detectors. We note that we find that our NIRSpec projections are weaker by a factor $\sim$2.

%%%%%%%%%%%%%%%%%%%%%%%%%%%%%%%%%%%%%%%%%%%%%%%%%%%%%%%%%%%%%%%%%%%%%%%%%%%

\section{Acknowledgements}
We thank Dylan Folsom, Joshua Foster, Mariia Khelashvili, Mariangela Lisanti, Hongwan Liu, Benjamin Safdi, and the JWST Space Telescope Science Institute help staff. A.P. acknowledges support from the Princeton Center for Theoretical Science postdoctoral fellowship. The work of C.B.~was supported in part by NASA through the NASA Hubble Fellowship Program grant HST-HF2-51451.001-A awarded by the Space Telescope Science Institute, which is operated by the Association of Universities for Research in Astronomy, Inc., for NASA, under contract NAS5-26555. Part of this work was done at the Aspen Center for Physics, which is supported by National Science Foundation grant PHY-1607611. S.R. was supported by the Department of Energy (DOE) under Award Number DE-SC0007968.

This research made extensive use of the publicly available codes \texttt{astropy}~\cite{astropy:2013,astropy:2018,astropy:2022}, \texttt{dust\_extinction}~\cite{dust_extinction}, \texttt{dustmaps}~\cite{Green_2019}, \texttt{HEALPix}~\cite{2005ApJ...622..759G},
\texttt{healpy}~\cite{Zonca2019},
\texttt{IPython}~\citep{PER-GRA:2007},
\texttt{Jupyter}~\citep{Kluyver2016jupyter},
\texttt{Mathematica}~\cite{Mathematica},
\texttt{matplotlib}~\citep{Hunter:2007},
\texttt{NumPy}~\citep{harris2020array}, \texttt{Python}~\cite{python3}
\texttt{SciPy}~\citep{2020SciPy-NMeth} and
\texttt{unyt}~\citep{Goldbaum2018}.

\bibliography{main.bib}

\clearpage

\onecolumngrid
\begin{center}
  \textbf{\large Supplemental Material for Sensitivity of JWST to eV-Scale Decaying Axion Dark Matter}\\[.2cm]
  \vspace{0.05in}
  {Sandip Roy, Carlos Blanco, Christopher Dessert, Anirudh Prabhu, Tea Temim}
\end{center}

\twocolumngrid

%%%%%%%%%% Merge with supplemental materials %%%%%%%%%%
\setcounter{equation}{0}
\setcounter{figure}{0}
\setcounter{table}{0}
\setcounter{section}{0}
\setcounter{page}{1}
\makeatletter
\renewcommand{\theequation}{S\arabic{equation}}
\renewcommand{\thefigure}{S\arabic{figure}}
\renewcommand{\theHfigure}{S\arabic{figure}}%
\renewcommand{\thetable}{S\arabic{table}}

\onecolumngrid

\section{Doppler Shifting and Broadening}
\label{app:Doppler}

\begin{figure}[h!]
\includegraphics[trim = {0cm, .5cm, 0cm, 0cm},width=0.5\textwidth]{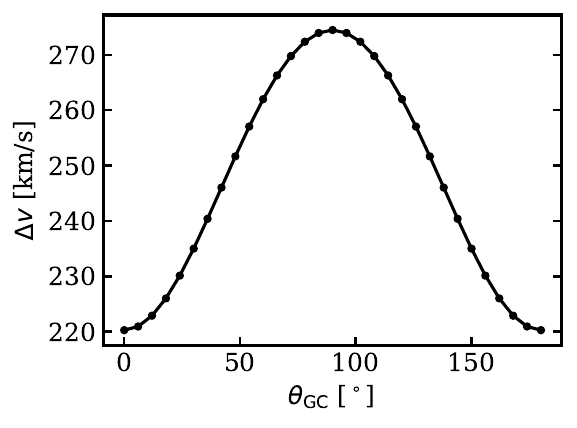}
\caption{
\label{fig:doppler_broadening}
Expected width of the stacked axion decay signal in each analysis annulus, including both Doppler shifting and broadening effects.
}
\end{figure}

In this section we describe our accounting for Doppler shifting and broadening in the ETC. We closely follow the approach of~\cite{Dessert:2023vyl}, which we do not repeat here. We assume a MW DM velocity dispersion of 220 km/s throughout the entire MW. However, the decay line is also Doppler shifted due to the solar motion relative to the DM rest frame. In galactic Cartesian coordinates, we have $\textbf{v}_\odot = (11,232,7)$ km/s~\cite{2000AA...354..522M,Sch_nrich_2010}. In our analysis we stack data within each galactocentric annulus, so the quantity of interest is the root-mean-squared line broadening due to solar Doppler shifting, given as $\sigma_\odot = (\int_\Omega d\Omega |\hat{\textbf{n}}\cdot\textbf{v}_\odot|^2/\int_\Omega d\Omega)^{1/2}$, where $\Omega$ is the area on the sky within the annulus. This is then added in quadrature with the DM velocity dispersion, so that the line width we enter into the ETC is given by $\Delta v = ((220\ \textrm{km/s})^2 + \sigma_\odot^2)^{1/2}$. In Fig.~\ref{fig:doppler_broadening}, we show $\Delta v$ in each of our annuli.

\section{Astrophysical Backgrounds}
\label{app:Backgrounds}
\newcommand{\phiGC}{\phi_{\rm GC}}

\begin{figure}[ht]
\centering
\includegraphics[trim = {0cm, .5cm, 0cm, 0cm},width=0.5\textwidth]{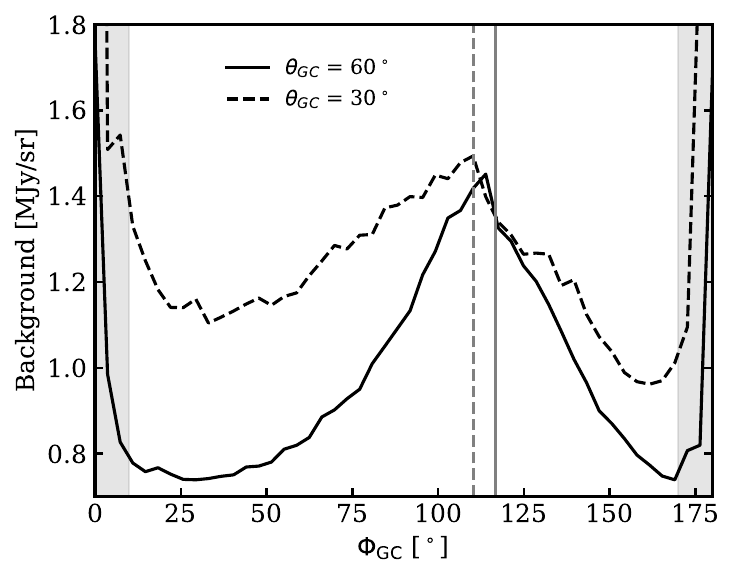}
\caption{Expected JWST background flux at $5\um$ at $\thetaGC = 30^\circ$ and $60^\circ$ as a function of the galactic azimuthal coordinate, $\phi_{\rm GC}$. Coordinates that intersect the ecliptic plane, which hosts larger backgrounds due to zodiacal emission, are marked in a gray vertical line. We shade the masked regions ($|\phiGC| \leq 10^\circ$ and $170^\circ \leq \phiGC \leq 190^\circ$). In our analysis, we conservatively assume the background flux at the ecliptic plane holds over the whole ring. Note that $0^\circ \leq\phiGC<360^\circ$, but we only show $\phiGC\leq180^\circ$ because the background is largely symmetric.
}
\label{fig:background_vs_phigal}
\end{figure}

We use the \texttt{jwst\_backgrounds} tool to compute the background flux at every point along each annular ring, labeled by the azimuthal coordinate $\phiGC$. In Fig.~\ref{fig:background_vs_phigal} we show the expected background at $5\um$ at $\thetaGC = 30^\circ$ and $60^\circ$ as a function of $\phiGC$. Except for the masked GP regions, the background flux is maximized when the ring intersects the ecliptic plane. The ETC takes in a coordinate at which to evaluate the background flux. In our projections, we conservatively enter the coordinate which maximizes the background over the ring.

\begin{figure}[ht]
\centering
\includegraphics[trim = {0cm, .5cm, 0cm, 0cm},width=0.49\textwidth]{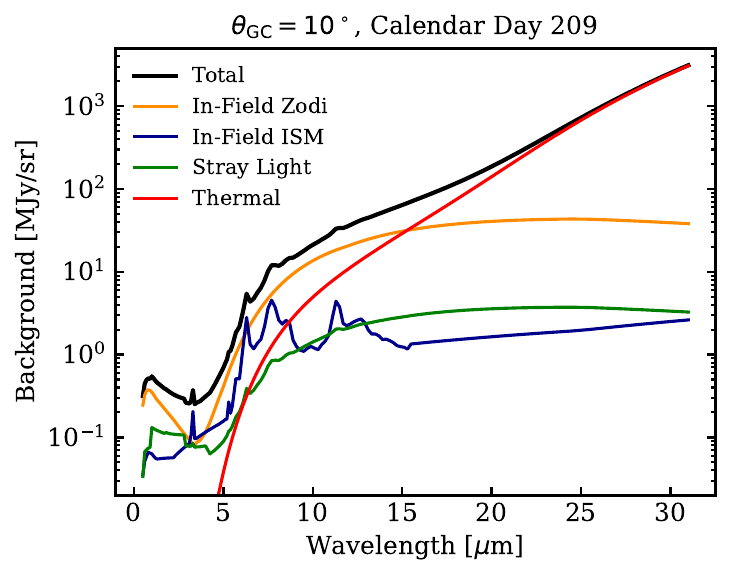}
\hfill % optional, for better spacing
\includegraphics[trim = {0cm, .5cm, 0cm, 0cm},width=0.49\textwidth]{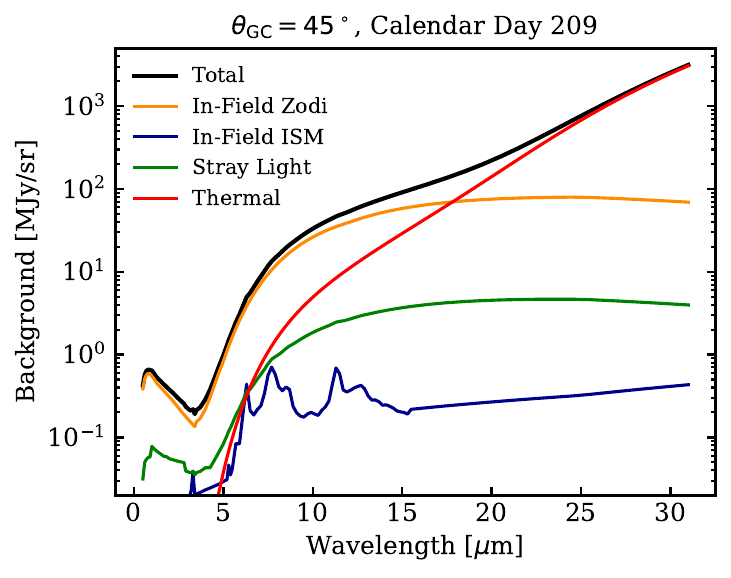}
\caption{Background components for $\thetaGC = 10^\circ$ and $45^\circ$ and on the ecliptic plane, generated using the \texttt{jwst\_backgrounds} tool. 
}
\label{fig:background_components}
\end{figure}

We show the wavelength-dependence of the JWST backgrounds in Fig.~\ref{fig:background_components} at the point where the annuli intersect with the ecliptic plane. The JWST background has four components: zodiacal (\ie solar system) emission, interstellar medium (ISM) emission, stray light which is scattered by the instrument from outside the FOV into the detector, and thermal emission of the instrument itself. At the lower galactic latitude, the in-field ISM emission is comparable to the in-field zodiacal emission at wavelengths $\lesssim 5\um$. At the greater galactic latitude, zodiacal emission dominates ISM emission. Thermal self-emission dominates at wavelengths $\gtrsim 15\um$ in both cases. We sum these background components to compute the background flux.

\newpage
\section{Dust Extinction}
\label{appendix:Extinction}

In this section we describe our treatment of the extinction due to dust in the MW. Electromagnetic radiation, particularly in the visual and UV bands, is scattered and absorbed on dust particles in the MW, increasing the opacity of the galaxy (for a review see~\cite{Draine:2003if}). In the near-IR the dust absorption cross-section is much smaller because the wavelength of the radiation is larger than the typical size of the dust particles. Although dust extinction is important for us, it does not substantially affect our results.

\begin{figure}[h!]
\includegraphics[trim = {0cm, 0.5cm, 0cm, 0cm},width=0.5\textwidth]{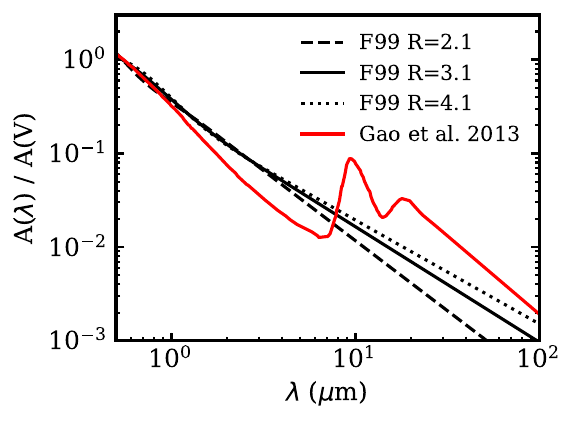}
\caption{
\label{fig:extinction_curves}
Dust extinction curves for different benchmark models from \cite{1998ApJ...500..525S,Gao_2013} which we extrapolate logarithmically to cover the wavelength range $0.5\um\,-\,100\um$. We use the red curve from \cite{Gao_2013}, calibrated specifically on the GC, for our analysis.
}
\end{figure}

\begin{figure}
\includegraphics[trim = {0cm, 0cm, 0cm, 0cm},width=0.49\textwidth]{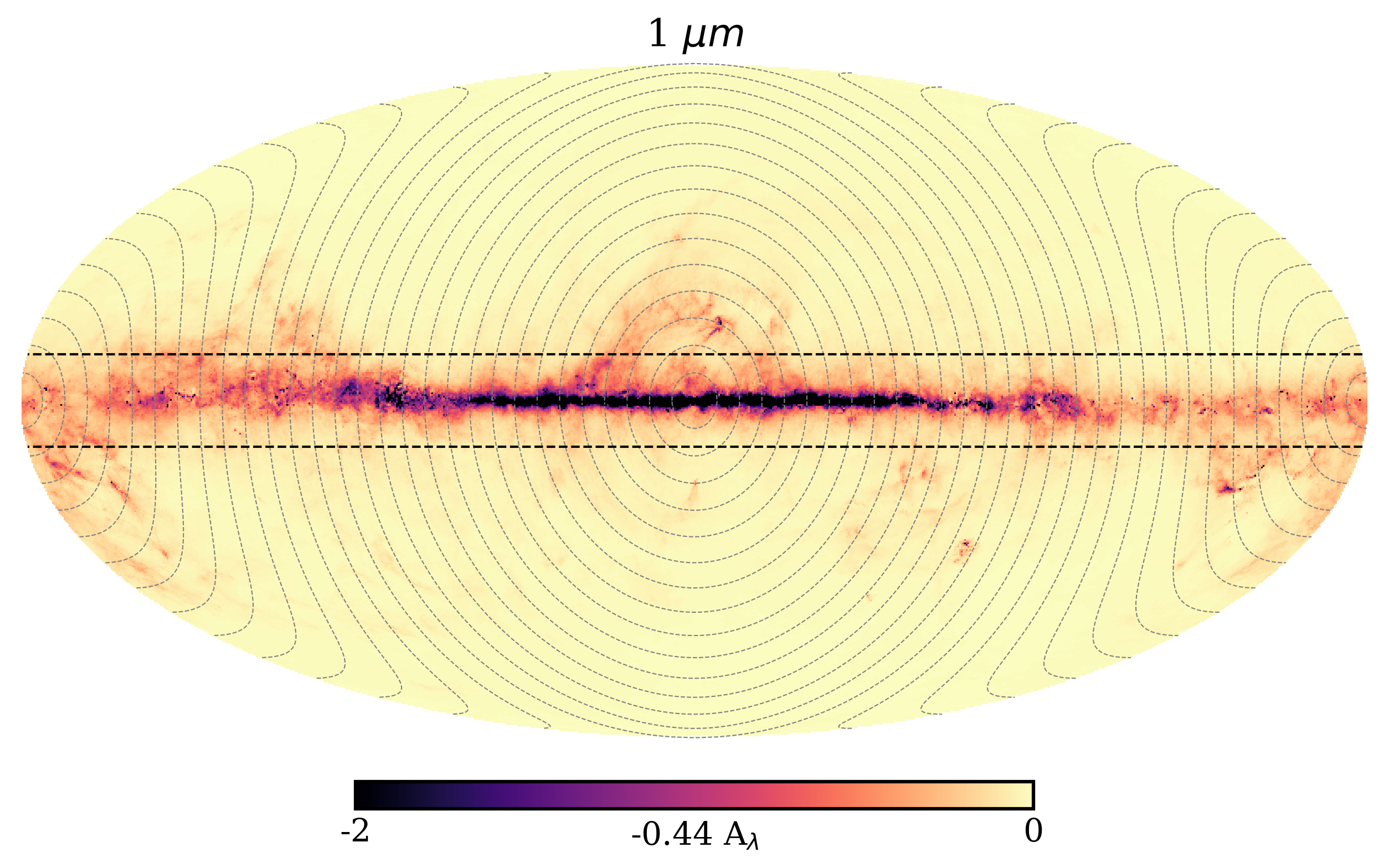}
\includegraphics[trim = {0cm, 0cm, 0cm, 0cm},width=0.49\textwidth]{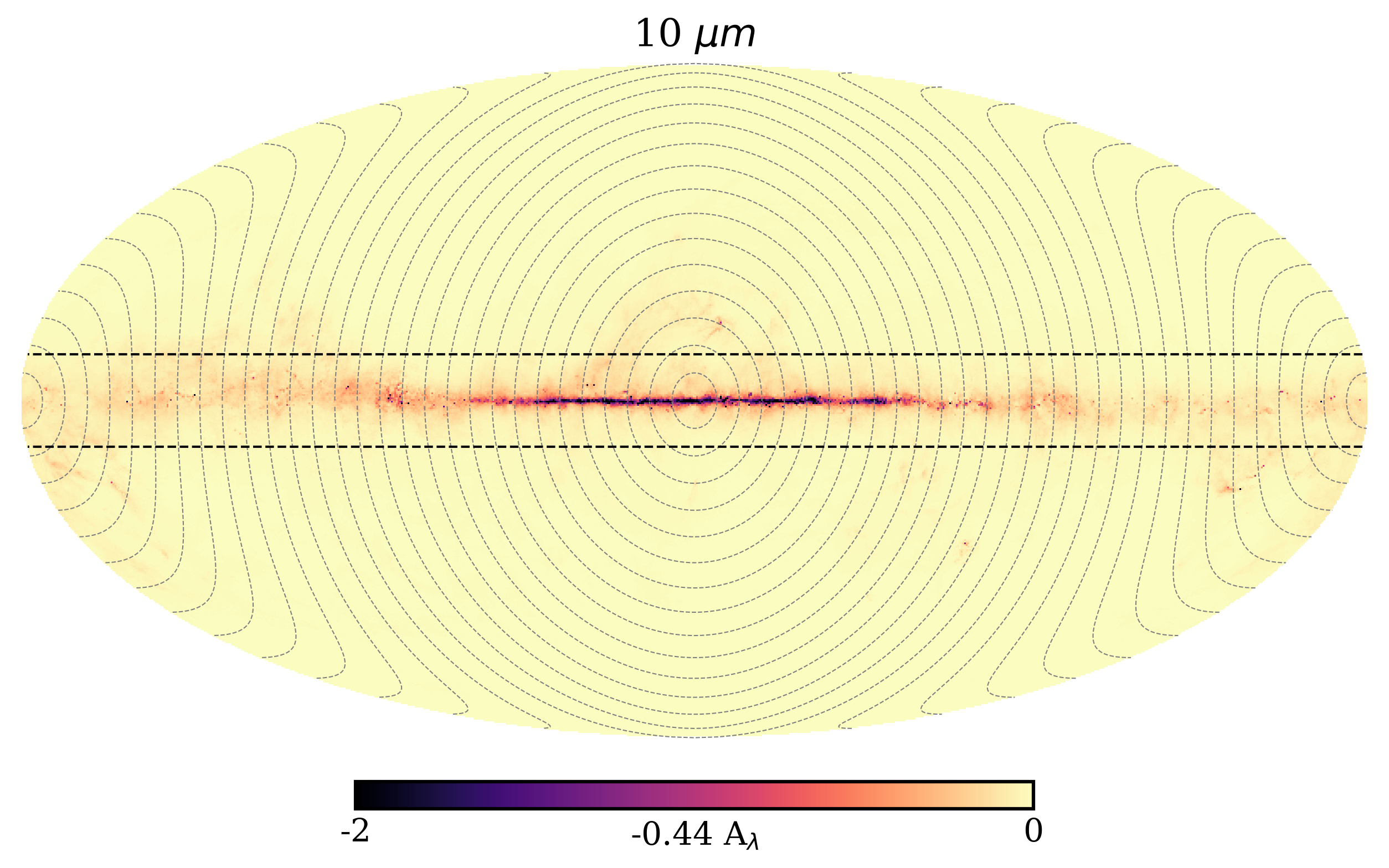}

\caption{
\label{fig:attenuation_maps}
\emph{Left panel}: Dust extinction map at $1\um$, corresponding to $m_a = 2.5\eV$, based on the dust models of \cite{1998ApJ...500..525S,doug_2011} and the extinction curve of \cite{Gao_2013}. The grey dotted lines delineate our ring edges and the dotted black lines are $|b|=10^\circ$, between which is masked in our analysis, similar to the scheme in Fig. \ref{fig:D-effective}.
\emph{Right panel}: Same as left panel but at $10\um$, corresponding to $m_a=0.25\eV$. As the mass of the axion decreases, the dust extinction becomes less significant according to the extinction law in Fig. \ref{fig:extinction_curves}. 
}
\end{figure}

\begin{figure}[h!]
\includegraphics[trim = {.5cm, 0.5cm, 0cm, 0cm},width=0.9\textwidth]{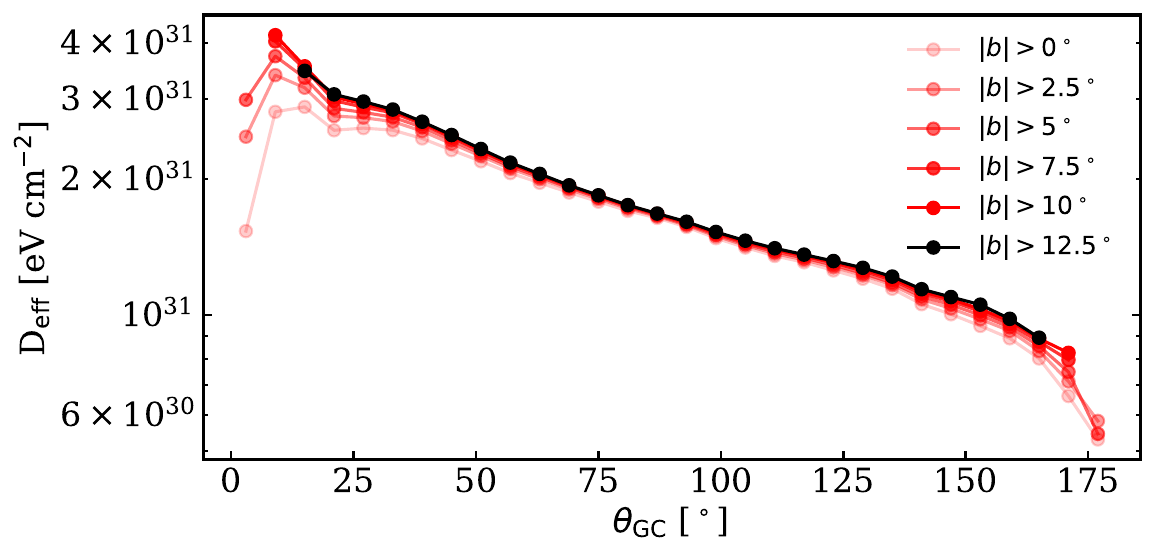}
\caption{
\label{fig:Deff_dust_exploration}
Line-of-sight averaged $\Deff$ factors for different galactic latitude ($|b|$) masks. As we increase the $|b|$ mask from $0^\circ$ to $10^\circ$, the mean $\Deff$ values near the GP experience less dust extinction. A too aggressive mask results in a loss of sensitivity because we exclude regions of large D-factor, as shown by the black line. Therefore we mask $|b|<10^\circ$ in our fiducial analysis.
}
\end{figure}

Electromagnetic radiation of wavelength $\lambda$ produced at a distance $d$ from Earth and galactic coordinates $(l,b)$ is reduced by a factor $\exp(-\int_0^d ds n_d(s,l,b) \sigma(\lambda))$. In the limit that the emission is produced outside of the galaxy, this expression vastly simplifies to $10^{-0.44 A_\lambda}$, where $A_\lambda$ is known as the galactic extinction (in units of magnitudes). We work in this limit throughout this manuscript, although we note that this is a conservative assumption as the DM decays throughout the galaxy. $A_\lambda$ varies both with sky coordinates and wavelength. The spatial morphology of the extinction is mapped out and computed in~\cite{1998ApJ...500..525S,doug_2011}, which provide $A_V$, the V-band extinction. We then use known scaling relations to compute the extinction at the relevant wavelengths, reproduced in Fig.~\ref{fig:extinction_curves}. The black lines correspond to models from~\cite{1998ApJ...500..525S} which are the generic expectations in the MW. In this work we use~\cite{Gao_2013}, which is the scaling relation derived using data at the Galactic Center. The dust extinction maps at $1\um$ and $10\um$ are shown in Fig.~\ref{fig:attenuation_maps}. In each analysis annulus, we compute the average effective D-factor $D_{\rm eff} = \int_\Omega d\Omega\,  D(l,b)\, 10^{-0.44 A_\lambda(l,b)}$, which we show in Fig.~\ref{fig:Deff_dust_exploration} for a variety of GP cuts.

\end{document}